\documentclass{WileyMSP-template}

\usepackage[numbers,sort&compress]{natbib}
\usepackage{amssymb}
\usepackage{tabularx}
\usepackage{color}
\newcolumntype{Y}{>{\centering\arraybackslash}X}

\begin{document}

\pagestyle{fancy}
\rhead{\includegraphics[width=2.5cm]{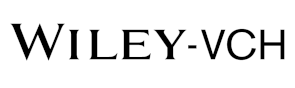}}

\title{Giant Magnetocaloric Effect in a Honeycomb Spiral Spin-Liquid Candidate}

\maketitle


\author{Yuqian Zhao,}
\author{Xun Chen,}
\author{Zongtang Wan,}
\author{Zhaohua Ma, and}
\author{Yuesheng Li*}

\dedication{}

\begin{affiliations}
Wuhan National High Magnetic Field Center and School of Physics, Huazhong University of Science and Technology, 430074 Wuhan, China\\
Email Address: \emph{*}yuesheng\_li@hust.edu.cn

\end{affiliations}


\keywords{magnetocaloric effect,  spiral spin liquid, frustrated honeycomb-lattice antiferromagnet, adiabatic demagnetization refrigeration}

\begin{abstract}
\begin{justify}
Unlike conventional magnetic states, which lack degeneracy, the spiral spin liquid (SSL) fluctuates among degenerate spiral configurations, with ground-state wave vectors forming a continuous contour or surface in reciprocal space. At low temperatures, the field-induced crossover from the polarized ferromagnetic state to the SSL results in a large entropy increase and decalescence, indicating its potential for magnetic cooling. However, magnetic cooling using a SSL has yet to be reported. Here, we investigate the magnetocaloric effect and cooling performance of single-crystal GdZnPO, a spin-7/2 honeycomb-lattice SSL candidate, under a magnetic field $H$ $<$ $H_\mathrm{c}$ ($\mu_0H_\mathrm{c}$ $\sim$ 12 T) applied perpendicular to the honeycomb plane and below the crossover temperature ($\sim$2 K). For $H$ $\geq$ $H_\mathrm{c}$, GdZnPO enters a polarized non-degenerate ferromagnetic state. Our results demonstrate that GdZnPO exhibits a giant low-temperature magnetocaloric effect near $H_\mathrm{c}$, surpassing other magnetocaloric materials. This giant magnetocaloric effect is well-explained by the frustrated honeycomb spin model of GdZnPO, suggesting the stability of the SSL below $H_\mathrm{c}$ down to very low temperatures. Additionally, its magnetic cooling performance remains robust up to at least 4.5 K, making GdZnPO a promising candidate for magnetic refrigeration down to $\sim$36 mK through cycling the applied magnetic field within a narrow range.
\end{justify}
\end{abstract}


\section{Introduction}

\begin{justify}
Paramagnetic materials with a large magnetocaloric effect (MCE) were initially proposed as a means to achieve temperatures well below 1 K, and have since developed into  an effective low-temperature refrigeration technique~\cite{debye1926einige,giauque1927thermodynamic}. Compared to conventional $^3$He-based refrigeration techniques (such as $^3$He-$^4$He dilution refrigerators), magnetic cooling is more cost-efficient, compact, and easier to handle~\cite{bartlett2015performance,bartlett2010improved,shirron2004portable,zheng2024optimization,gruner2024metallic}, as $^3$He gas is scarce, non-renewable, and expensive~\cite{jang2015large}. In the pursuit of higher MCE and improved efficiency in magnetic cooling, a novel approach based on a field-induced critical point (CP) in strongly correlated and frustrated spin systems has been proposed and successfully tested~\cite{PhysRevB.67.104421,wolf2011magnetocaloric,wolf2016magnetic,xiang2024giant}. This technique offers at least three key advantages: (1) Strongly correlated magnets typically have a larger volumetric entropy capacity, $S_\mathrm{V}$ = $R$ln(2$s$+1)/$V_\mathrm{m}$, compared to highly diluted paramagnetic salts, enabling further reduction in magnet size and ensuring good cooling power. Here, $V_\mathrm{m}$ is the volume per mole of spins and $s$ is the spin quantum number. (2) The MCE of strongly correlated magnets can be significantly enhanced, with a low-temperature magnetic Gr\"{u}neisen parameter $\it{\Gamma}_\mathrm{m}$ = $(\mathrm{d}T/\mathrm{d}H)/(\mu_0T)$ $\sim$ 0.6-3.8 T$^{-1}$~\cite{PhysRevX.10.011007,wolf2011magnetocaloric,wolf2016magnetic,xiang2024giant}, as the applied magnetic field approaches the critical value. (3) During the refrigeration cycles, the magnetic field change can be confined to a relatively narrow range, where $\it{\Gamma}_\mathrm{m}$ is large, reducing eddy current effects in metal parts and minimizing energy cost.

A spiral spin liquid (SSL) in a frustrated spin system fluctuates cooperatively among degenerate spiral configurations, with ground-state wave vectors ($\mathbf{Q}_\mathrm{G}$) forming a continuous contour or surface in reciprocal space, depending on the system~\cite{bergman2007order,yao2021generic,PhysRevResearch.4.023175}. The low-energy topological excitations and defects in a SSL, such as spin vortices, antivortices~\cite{PhysRevB.93.085132,gao2017spiral}, and momentum vortices~\cite{PhysRevResearch.4.023175,PhysRevB.110.085106}, may be used in small-scale antiferromagnetic spintronic devices~\cite{jungwirth2016antiferromagnetic,RevModPhys.90.015005,gao2020fractional}, offering topologically protected memory and logic operations~\cite{yao2013topologically} without magnetic field leakage. Because of the degenerate spiral contour (see Figure~\ref{fig1}a) or surface, the specific heat of a SSL ($C_\mathrm{SSL}$) remains large even at extremely low temperatures, $C_\mathrm{SSL}$($T$ $\to$ 0) $\to$ $C_0$ ($\sim$$R/$2 for a two-dimensional classical SSL)~\cite{bergman2007order,yao2021generic,PhysRevLett.133.236704}, in contrast to a conventional magnetic ordered phase (CMOP, see Figure~\ref{fig1}b), where $C_\mathrm{CMOP}$($T$ $\to$ 0) $\to$ 0. As a result, the entropy ($S$) of the SSL is significantly higher than that of the conventional magnetic ordered phase at low temperatures. The crossover from the conventional magnetic ordered phase to the SSL in a frustrated spin system produces large decalescence, $Q$ = $\int_\mathrm{CMOP}^\mathrm{SSL}TdS$, highlighting the potential of SSL candidates for magnetic cooling near the CP. However, to the best of our knowledge, magnetic cooling using a SSL has not yet been explored.

Under a magnetic field applied perpendicular to the plane, SSLs can exist over a wide range of interaction parameters in the easy-plane ($D$ $\geq$ 0) frustrated honeycomb-lattice model: $\mathcal{H} = J_1\sum_{\langle j0,j1\rangle}\mathbf{s}_{j0}\cdot\mathbf{s}_{j1}+J_2\sum_{\langle\langle j0,j2\rangle\rangle}\mathbf{s}_{j0}\cdot\mathbf{s}_{j2}+D\sum_{j0}(s_{j0}^z)^2-\mu_0Hg\mu_\mathrm{B}\sum_{j0}s_{j0}^z$, where $J_1$ and $J_2$ are the first- and second-nearest-neighbor Heisenberg couplings, respectively, and $\mu_\mathrm{B}$ is Bohr magneton~\cite{PhysRevB.81.214419,owerre2017topological,PhysRevB.106.035113,okumura2010novel,PhysRevB.100.224404,PhysRevResearch.4.013121}. In this model, $\mathbf{Q}_\mathrm{G}$ forms a continuous contour around the $\Gamma\{0, 0\}$ point (1/2 $>$ $|J_2/J_1|$ $>$ 1/6) or the K$\{1/3, 1/3\}$ point ($|J_2/J_1|$ $>$ 1/2) in reciprocal space. In the previous work~\cite{PhysRevLett.133.236704}, we identified the structurally disorder-free rare-earth Gd$^{3+}$ ($s$ = 7/2) antiferromagnet GdZnPO~\cite{nientiedt1998equiatomic,lincke2008magnetic} as a candidate for this prototypical honeycomb-lattice model, with $J_1$ $\sim$ $-$0.39 K, $J_2$ $\sim$ 0.57 K, $D$ $\sim$ 0.30 K, and the $g$ factor $g$ $\sim$ 2. With $|J_2/J_1|$ $\sim$ 1.5 ($>$ 1/2), GdZnPO undergoes a crossover to a unique SSL with a spiral contour around the K point below $T^*$ $\sim$ 2 K, from a thermally paramagnetic phase above $T^*$~\cite{PhysRevLett.133.236704}. The SSL contour is analytically given by $\mathbf{Q}_\mathrm{G}$ = $h_\mathrm{G}\mathbf{b}_1$+$k_\mathrm{G}\mathbf{b}_2$ with the condition $\cos(2\pi h_\mathrm{G})+\cos(2\pi k_\mathrm{G})+\cos(2\pi h_\mathrm{G}+2\pi k_\mathrm{G}) = [J_1^2/(4J_2^2)-3]/2$ (Figure~\ref{fig1}a), which is independent of $H$ below the crossover value $H_\mathrm{c}$. Here, $\mathbf{b}_1$ and $\mathbf{b}_2$ are the reciprocal lattice vectors.

\begin{figure}[h!]
\begin{center}
  \includegraphics[width=10cm,angle=0]{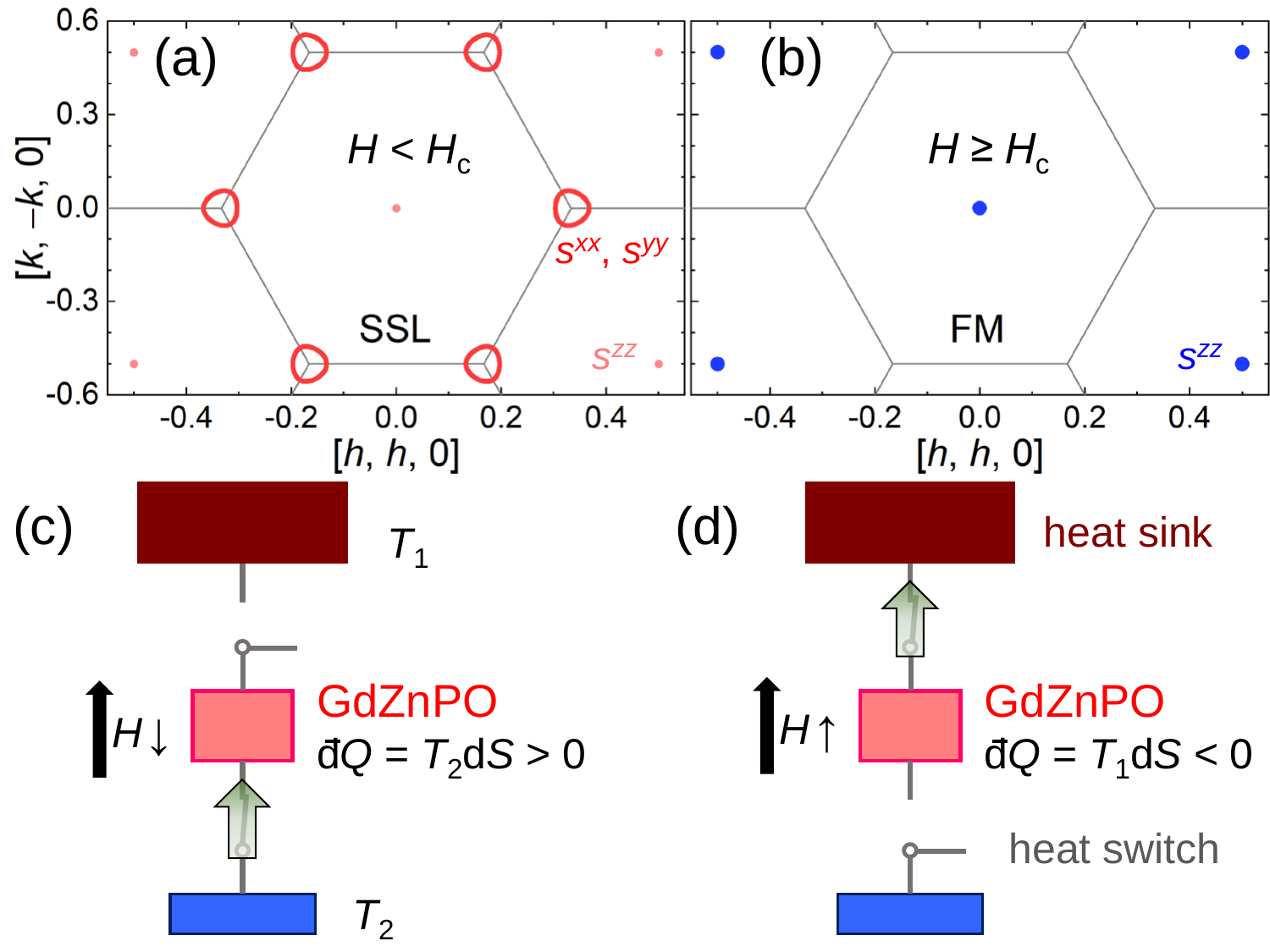}
  \begin{justify}
  \caption{Magnetic cooling using GdZnPO. a,b) Ground-state wave vectors ($\mathbf{Q}_\mathrm{G}$) of the spiral spin liquid (SSL) and the polarized ferromagnetic (FM) phase, with magnetic fields applied along the $c$ axis, for $H$ $<$ $H_\mathrm{c}$ and $H$ $\geq$ $H_\mathrm{c}$, respectively. The in-plane spin structure factor $s^{xx}$ or $s^{yy}$ forms a degenerate $\mathbf{Q}_\mathrm{G}$ contour around the K$\{1/3, 1/3\}$ point in the SSL, while the out-of-plane component $s^{zz}$ is nonzero only at the $\Gamma\{0, 0\}$ point for $H$ $\neq$ 0~\cite{PhysRevLett.133.236704}. $H_\mathrm{c}$ is the crossover field, and the gray lines represent Brillouin zone boundaries. c,d) Schematic diagrams of heat absorption ($\mathrm{d}\kern-3.0pt\bar{~}Q$ $>$ 0) and release ($\mathrm{d}\kern-3.0pt\bar{~}Q$ $<$ 0) processes due to entropy changes.}
  \label{fig1}
  \end{justify}
\end{center}
\end{figure}

Below $T^*$, the GdZnPO spin system stays in the putative SSL with the degenerate contour unchanged for $H$ $<$ $H_\mathrm{c}$, but crosses over to a substantially polarized ferromagnetic (FM) phase when $H$ $\geq$ $H_\mathrm{c}$~\cite{PhysRevLett.133.236704}. The crossover magnetic field is analytically given by $\mu_0H_\mathrm{c}$ = $s[2D+3J_1+9J_2+J_1^2/(4J_2)]/(g\mu_\mathrm{B})$ ($\sim$12 T). Because of the spatial localization of the 4$f$ Gd$^{3+}$ electrons, both $J_1$ and $J_2$ interactions are relatively weak, making the crossover field ($\mu_0H_\mathrm{c}$ $\sim$ 12 T) experimentally accessible. Moreover, single-crystal samples are available, enabling the magnetic field to be applied perpendicular to the honeycomb plane, i.e., along the $c$ axis. Therefore, single-crystal GdZnPO is an ideal candidate for magnetic cooling near $H_\mathrm{c}$.

In this work, we investigate the low-temperature MCE of single-crystal GdZnPO with the magnetic field applied along the $c$ axis to assess its magnetic cooling capacity. GdZnPO, with a large $S_\mathrm{V}$ = 0.424 JK$^{-1}$cm$^{-3}$, exhibits a giant MCE with a normalized Gr\"{u}neisen ratio $\it{\Gamma}^\mathrm{norm}$ = $\mu_0H\it{\Gamma}_\mathrm{m}$ = 28-51 between 11 and 12 T (near the crossover field $\mu_0H_\mathrm{c}$ $\sim$ 12 T) at low temperatures, where $\it{\Gamma}_\mathrm{m}$ = 2.5-4.4 T$^{-1}$ is the magnetic Gr\"{u}neisen parameter. Both $\it{\Gamma}^\mathrm{norm}$ and magnetic cooling performance remain high up to at least 4.5 K ($>$ $T^*$), suggesting the promising potential of GdZnPO, a honeycomb-lattice SSL candidate, for adiabatic demagnetization refrigeration. These results may encourage further research into the use of other SSL candidates~\cite{bergman2007order,gao2017spiral,PhysRevB.98.064427,gao2020fractional,PhysRevLett.130.166703,gao2024,PhysRevB.103.104433,PhysRevB.104.024426,takahashi2024,PhysRevLett.128.227201} for magnetic cooling.
\end{justify}

\section{Results}

\subsection{Magnetocaloric effect}

\begin{figure}[h!]
\begin{center}
  \includegraphics[width=11cm,angle=0]{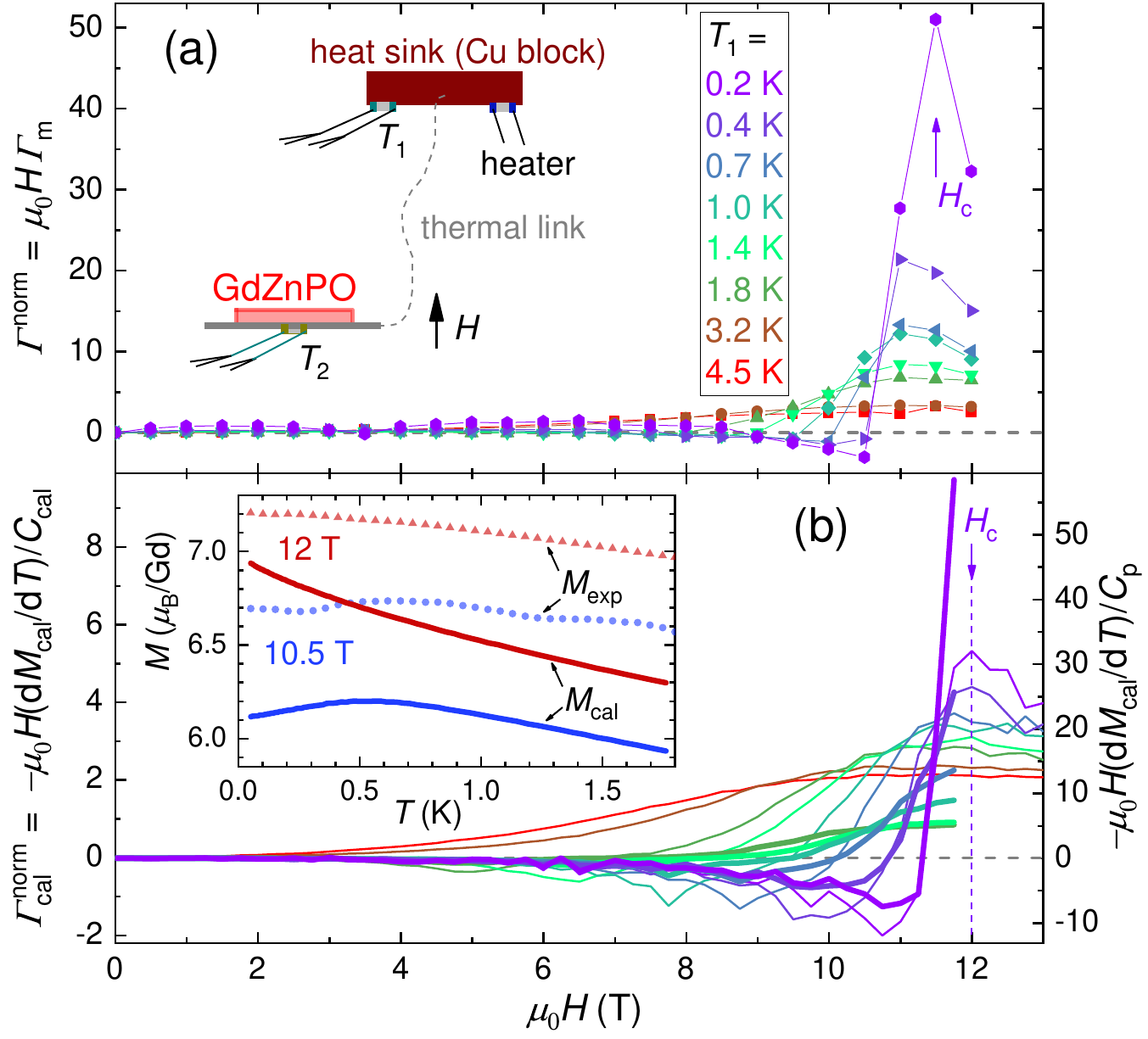}
  \begin{justify}
  \caption{Magnetocaloric effect of GdZnPO. a) Magnetic field dependence of the normalized Gr\"{u}neisen ratio, $\it{\Gamma}^\mathrm{norm}$ = $\mu_0H\it{\Gamma}_\mathrm{m}$, measured with an alternating-field frequency of 0.005 Hz, where $\it{\Gamma}_\mathrm{m}$ is the magnetic Gr\"{u}neisen parameter. Inset: Diagrammatic sketch of the experimental setup. b) Gr\"{u}neisen ratio calculated from Monte Carlo simulations of magnetization ($M_\mathrm{cal}$) and specific heat ($C_\mathrm{cal}$), $\it{\Gamma}^\mathrm{norm}_\mathrm{cal}$ = $-\mu_0H(\mathrm{d}M_\mathrm{cal}/\mathrm{d}T)/C_\mathrm{cal}$ (thin lines). Thick lines show $-\mu_0H(\mathrm{d}M_\mathrm{cal}/\mathrm{d}T)/C_\mathrm{p}$, using experimental specific heat $C_\mathrm{p}$ instead. Inset: Temperature dependence of $M_\mathrm{cal}$ (lines) and experimental magnetization $M_\mathrm{exp}$ (points) at 10.5 and 12 T. The crossover field $H_\mathrm{c}$ is indicated.}
  \label{fig2}
  \end{justify}
\end{center}
\end{figure}

\begin{justify}
The MCE characterizes the magnetic cooling performance of candidate compounds. In our measurements on GdZnPO, the alternating magnetic field is generated by the main superconducting magnet as $H(t)$ = $\Delta H\sin[2\pi\nu(t-t_H)]+H^\mathrm{av}$, which induces a sine wave signal for the sample temperature $T_2(t)$ $\sim$ $\Delta T\sin[2\pi\nu(t-t_T)]+T^\mathrm{av}$, with $\nu$ = 0.005 Hz. The normalized Gr\"{u}neisen ratio is given by $\it{\Gamma}^\mathrm{norm}$($T$ = $T^\mathrm{av}$ $\sim$ $T_1$, $H$ = $H^\mathrm{av}$) = $\Delta TH/(T\Delta H)$ (see Figure~\ref{fig2}a). The fits yield $t_T$ $\approx$ $t_H$ (Figure S1, Supporting Information), indicating good thermal contact between the GdZnPO crystals and the thermometer $T_2$~\cite{tokiwa2011high}.
\end{justify}

\begin{justify}
\quad Below 1.8 K, the maximum heat leak power due to the weak thermal link (see inset of Figure~\ref{fig2}a) is calculated as $P_\mathrm{L}^\mathrm{max}$ = $C_\mathrm{tot}|\Delta T|/\tau$, where the total specific heat $C_\mathrm{tot}$ is predominantly ($\gtrsim$ 99.8\%) contributed by the magnetic specific heat of the GdZnPO sample, and $\tau$ = 400-600 s is the relaxation time (see C$\to$A processes in Figure~\ref{fig3}b). At $T$ $\sim$ 1.8 K, the specific heat of GdZnPO and its nonmagnetic reference YZnPO are 9.8 and 0.0156 JK$^{-1}$mol$^{-1}$, respectively~\cite{PhysRevLett.133.236704}. As $T$ $\to$ 0 K, the lattice specific heat decreases following a $T^3$ behavior, much more rapidly than the magnetic contribution. In contrast, the maximum cooling or heating power due to the alternating field is $P_\mathrm{C}^\mathrm{max}$ $\sim$ $2\pi\nu C_\mathrm{tot}|\Delta T|$. Since $\tau^{-1}$ $\ll$ $2\pi\nu$, it follows that $P_\mathrm{L}^\mathrm{max}$ $\ll$ $P_\mathrm{C}^\mathrm{max}$, confirming the quasi-adiabatic conditions of our experiments. Moreover, doubling the frequency to $\nu$ = 0.01 Hz does not significantly increase $\it{\Gamma}_\mathrm{m}$ or $\it{\Gamma}^\mathrm{norm}$ (Figure S2, Supporting Information), further validating our results.

The temperature and magnetic field dependence of the normalized Gr\"{u}neisen ratio is simulated by Monte Carlo (MC) calculations via $\it{\Gamma}^\mathrm{norm}_\mathrm{cal}$ = $-\mu_0H(\mathrm{d}M_\mathrm{cal}/\mathrm{d}T)/C_\mathrm{cal}$ (thin lines in Figure~\ref{fig2}b)~\cite{tokiwa2011high}. At low temperatures, $\it{\Gamma}^\mathrm{norm}_\mathrm{cal}$ peaks at the crossover field $\mu_0H_\mathrm{c}$ $\sim$ 12 T of the many-body model~\cite{PhysRevLett.133.236704}. Experimentally, $\it{\Gamma}^\mathrm{norm}$ also peaks at $\sim$11.5 T ($\sim$$\mu_0H_\mathrm{c}$), where the magnetization $M$ approaches its fully polarized value of $sg$ $\sim$ 7 $\mu_\mathrm{B}$/Gd, at low temperatures (Figure S2, Supporting Information). Therefore, the maximum field of our superconducting magnet, $\mu_0H$ = 12 T, is sufficient to substantially polarize the GdZnPO spin system. However, $\it{\Gamma}^\mathrm{norm}_\mathrm{cal}$ is smaller than the experimental $\it{\Gamma}^\mathrm{norm}$, near $H_\mathrm{c}$. This discrepancy is primarily due to the spin quantization ($s$ = 7/2) in the real GdZnPO compound. This quantization reduces the entropy per mole of spins to a finite value of $R$ln(2$s$+1), thereby strongly suppressing the specific heat at low temperatures. Replacing the overestimated specific heat from the classical model ($C_\mathrm{cal}$) with the experimentally measured value ($C_\mathrm{p}$) significantly reduces the discrepancy in the Gr\"{u}neisen ratio, as shown by the thick lines in Figure~\ref{fig2}b. The comparison between experimental and calculated specific heat was extensively discussed in our previous work on GdZnPO~\cite{PhysRevLett.133.236704}.

Moreover, the classical MC calculations yield a crossover field of $\mu_0H_\mathrm{c}$ = 12 T, which appears slightly higher than the experimental value. However, the low-$T$ d$M$/d$H$ remains larger than the Van Vleck susceptibility ($\chi_\mathrm{vv}^\parallel$ $\sim$ 0.3 cm$^3$mol$^{-1}$) even at 12 T (Figure S3, Supporting Information), indicating that the GdZnPO spin system is not yet fully polarized at this field. This discrepancy is also likely due to quantum fluctuations associated with the finite spin quantum number $s$ = 7/2 of the Gd$^{3+}$ ions, which are not captured by the classical model.

The magnetocaloric effect can be quantitatively evaluated using magnetization ($M$) and specific heat ($C_\mathrm{p}$) measurements, via $\it{\Gamma}_\mathrm{m}$ = $-(\mathrm{d}M/\mathrm{d}T)/C_\mathrm{p}$. The low-$T$ $C_\mathrm{m}$ ($\sim$ $C_\mathrm{p}$) exhibits a sharp drop as the applied magnetic field approaches to the crossover (or critical) value of $\mu_0H_\mathrm{c}$ $\sim$ 12 T (Figure S3, Supporting Information), which predominantly contributes to the observed giant low-temperature magnetocaloric effect. The Van Vleck magnetization is expected to be temperature-independent~\cite{PhysRevLett.133.236704} and can thus be safely neglected in simulations of the magnetocaloric effect. At low temperatures, a negative value of $\it{\Gamma}_\mathrm{m}$ implies that $\mathrm{d}M/\mathrm{d}T$ $>$ 0, meaning the magnetization $M$ increases with temperature $T$. For example, at 0.2 K and 10.5 T, we observe $\it{\Gamma}_\mathrm{m}$ $<$ 0 in both the experimental and simulated data (see Figure~\ref{fig2}), and correspondingly, $M$ increases with $T$ (see the inset of Figure~\ref{fig2}b). Negative $\it{\Gamma}_\mathrm{m}$ has also been widely reported at low temperatures in other frustrated magnets, such as $\alpha$-RuCl$_3$~\cite{PhysRevB.103.054440} and TmMgGaO$_4$~\cite{PhysRevX.10.011007}. In weakly correlated spin systems at finite temperatures, or strongly correlated ones at high temperatures, magnetization follows the Curie-Weiss law under the mean-field approximation, giving $\it{\Gamma}_\mathrm{m}$ $>$ 0. Thus, observing $\it{\Gamma}_\mathrm{m}$ $<$ 0 in a strongly correlated insulating magnet likely reflects complex many-body spin correlations emerging at low temperatures ($<$ $|\theta_\mathrm{w}|$). Here, $\theta_\mathrm{w}$ $\sim$ $-$12 K is the Curie-Weiss temperature determined from magnetic susceptibility measurements~\cite{PhysRevLett.133.236704}.

Below $\sim$$T^*$, as $\mu_0H$ decreases to $\sim$10.5 T, $\it{\Gamma}^\mathrm{norm}$ changes sign (Figure~\ref{fig2}a), as confirmed by the raw alternating-field data (Figure S1, Supporting Information). This feature is well reproduced by MC simulations, where $M_\mathrm{cal}$ increases at $\sim$12 T but decreases at $\sim$10.5 T as $T$ approaches 0 K, aligning with the noisy experimental magnetization $M_\mathrm{exp}$. $M_\mathrm{exp}$ exceeds $M_\mathrm{cal}$ due to the Van Vleck contribution (inset of Figure~\ref{fig2}b). As $T$ $\to$ 0 K, the magnetic field where $\it{\Gamma}^\mathrm{norm}$ = 0 approaches $H_\mathrm{c}$, where $\it{\Gamma}^\mathrm{norm}$ reaches a maximum (see Figure~\ref{fig2}), indicating the second-order nature of the field-induced crossover~\cite{PhysRevB.103.054440,PhysRevB.102.104433}. Additionally, the high-resolution magnetization measured at 50 mK shows no evident magnetic hysteresis (Figure S2, Supporting Information), confirming that the field-induced crossover is second-order or higher. The directly determined Gr\"{u}neisen parameter $\it{\Gamma}_\mathrm{m}$ is also roughly consistent with the noisy data obtained indirectly using experimentally measured $M$ and $C_\mathrm{p}$, where $\it{\Gamma}_\mathrm{m}$ $\sim$ $-(\mathrm{d}M/\mathrm{d}T)/C_\mathrm{p}$ (Figure S4, Supporting Information).

As the GdZnPO spin system becomes thermally paramagnetic above $T^*$~\cite{PhysRevLett.133.236704}, $\it{\Gamma}^\mathrm{norm}$ remains positive and moderately high ($\sim$3.3 at 4.5 K and 11.5 T) across the full range of $\sim$0 and 12 T (see Figure~\ref{fig2}a and Table~\ref{tab1}). In contrast, we highlight that the observed $\it{\Gamma}^\mathrm{norm}$ $\sim$ 51 ($\gg$ 3.3) well below $T^*$ and near the crossover field is exceptionally high, arising from the crossover between the degenerate SSL and the non-degenerate FM phases, which forms the core finding of this work. At $H$ $<$ $H_\mathrm{c}$ in the $J_1$-$J_2$ frustrated honeycomb-lattice model, the SSL is predicted to remain highly stable without ``order by disorder" down to very low temperatures~\cite{PhysRevResearch.4.013121}. This is confirmed by the giant magnetocaloric effect observed in this work and our previous thermodynamic measurements on GdZnPO~\cite{PhysRevLett.133.236704}, supporting its use in magnetic cooling down to the mK range (see below).
\end{justify}

\subsection{Quasi-adiabatic demagnetization refrigeration}

\begin{figure}[h!]
\begin{center}
  \includegraphics[width=11cm,angle=0]{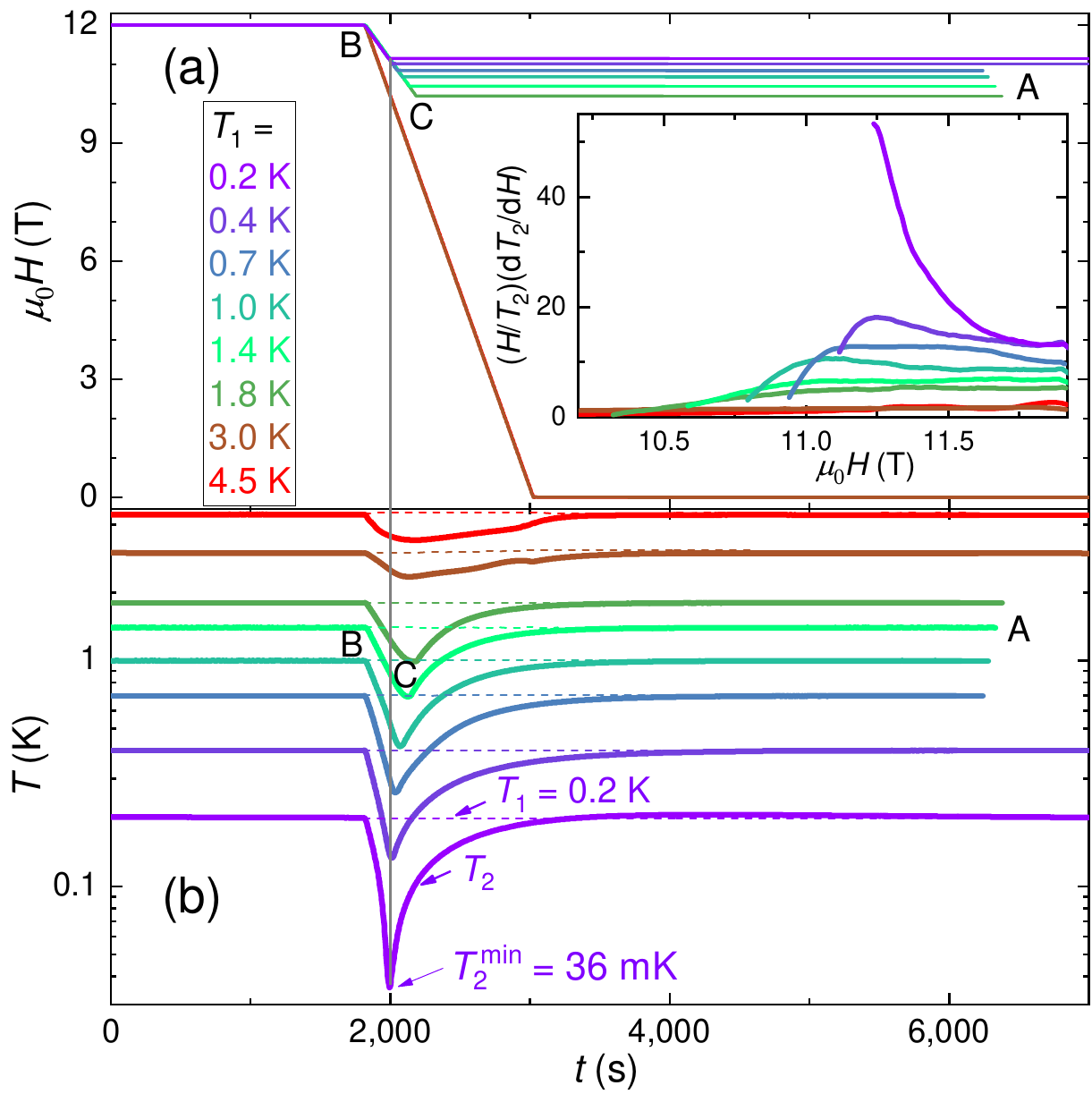}
  \begin{justify}
  \caption{Quasi-adiabatic demagnetization refrigeration using GdZnPO. a,b) Time dependence of the magnetic field and temperatures, $T_1$ (dashed colored lines) and $T_2$ (solid colored lines). The heat sink temperature, $T_1$, was controlled at various values using the heater shown in the inset of Figure~\ref{fig2}a. The inset of panel a) shows the corresponding Gr\"{u}neisen ratio, $(H/T_2)(\mathrm{d}T_2/\mathrm{d}H)$. A high field ramp rate, $v_H$ = 0.6 Tmin$^{-1}$, was used for $T_1$ $\geq$ 3 K, while $v_H$ = 0.3 Tmin$^{-1}$ was applied for $T_1$ $\leq$ 1.8 K to reduce eddy heat.}
  \label{fig3}
  \end{justify}
\end{center}
\end{figure}

\begin{justify}
In the narrow range between the starting field $\mu_0H_\mathrm{sta}$ $\sim$ 12 T and the end field $\mu_0H_\mathrm{end}$ $\sim$ 11 T, $\it{\Gamma}^\mathrm{norm}$ remains high across a range of temperatures 0.2-4.5 K (Figure~\ref{fig2}a), strongly indicating GdZnPO's potential for adiabatic demagnetization refrigeration. Using heat switches (HSs), adiabatic demagnetization refrigeration cycles can be implemented as illustrated in Figure~\ref{fig1}c,d. Several methods exist for implementing HSs~\cite{Duband1995,bartlett2010improved,ram2023critical,bartlett2015performance}; in this work, we used the above setup (inset of Figure~\ref{fig2}a) to perform the quasi-adiabatic demagnetization refrigeration (Figure~\ref{fig3}) due to the absence of HSs in our lab. By using a higher $H_\mathrm{sta}$, incorporating HSs, and employing a larger single-crystal sample, the magnetic cooling performance of GdZnPO during each run (Figure~\ref{fig3}) could be significantly improved under better adiabatic conditions. Since $\tau$ = $C_\mathrm{tot}$/$\kappa_\mathrm{eff}$, a larger single-crystal sample leads to a greater $C_\mathrm{tot}$, a longer relaxation time $\tau$, and thus better adiabatic conditions. Here, $\kappa_\mathrm{eff}$ represents the effective heat conductance between the sample and the heat sink.

When the applied field decreases from $H_\mathrm{sta}$ to $H_\mathrm{end}$, the GdZnPO spin system evolves from the polarized FM state to the degenerate SSL, absorbing heat due to the entropy increase during the quasi-adiabatic demagnetization refrigeration and the subsequent constant-$H_\mathrm{end}$ relaxation processes (B$\to$C$\to$A, see Figure~\ref{fig4}a). Figure~\ref{fig3} shows measurements during the processes B$\to$C$\to$A, with the heat sink temperature $T_1$ maintained at various values from 0.2 to 4.5 K (dashed lines in Figure~\ref{fig3}b). At $T_1$ = 0.2 K, the lowest sample temperature $T_2$$^\mathrm{min}$ = 36 mK is achieved by reducing the field from $\mu_0H_\mathrm{sta}$ = 12 T to $\mu_0H_\mathrm{end}$ = 11.2 T at a ramp rate of $v_H$ = 0.3 Tmin$^{-1}$, with good repeatability (Figure S2, Supporting Information). Our results demonstrate that efficient adiabatic demagnetization refrigeration using GdZnPO can be performed in a narrow field range between $\sim$11 and 12 T. Furthermore, the magnetic cooling performance of GdZnPO remains effective up to at least $T_1$ = 4.5 K, above the temperature of liquid $^4$He (4.2 K) and well above $T^*$. The normalized Gr\"{u}neisen ratio is obtained as $\it{\Gamma}^\mathrm{norm}$ $\sim$ $(H/T_2)(\mathrm{d}T_2/\mathrm{d}H)$, and is roughly consistent with the high-resolution value determined by the alternating-field technique (Figure~\ref{fig2}a). However, the Gr\"{u}neisen ratio shown in the inset of Figure~\ref{fig3}a was measured under non-isothermal conditions with a large $T_1-T_2$, making it less reliable. Here, $\it{\Gamma}^\mathrm{norm}$ was underestimated due to the heat leak power, which is proportional to $T_1-T_2$.
\end{justify}

\subsection{Magnetic cooling power and efficiency}

\begin{figure}[h!]
\begin{center}
  \includegraphics[width=13cm,angle=0]{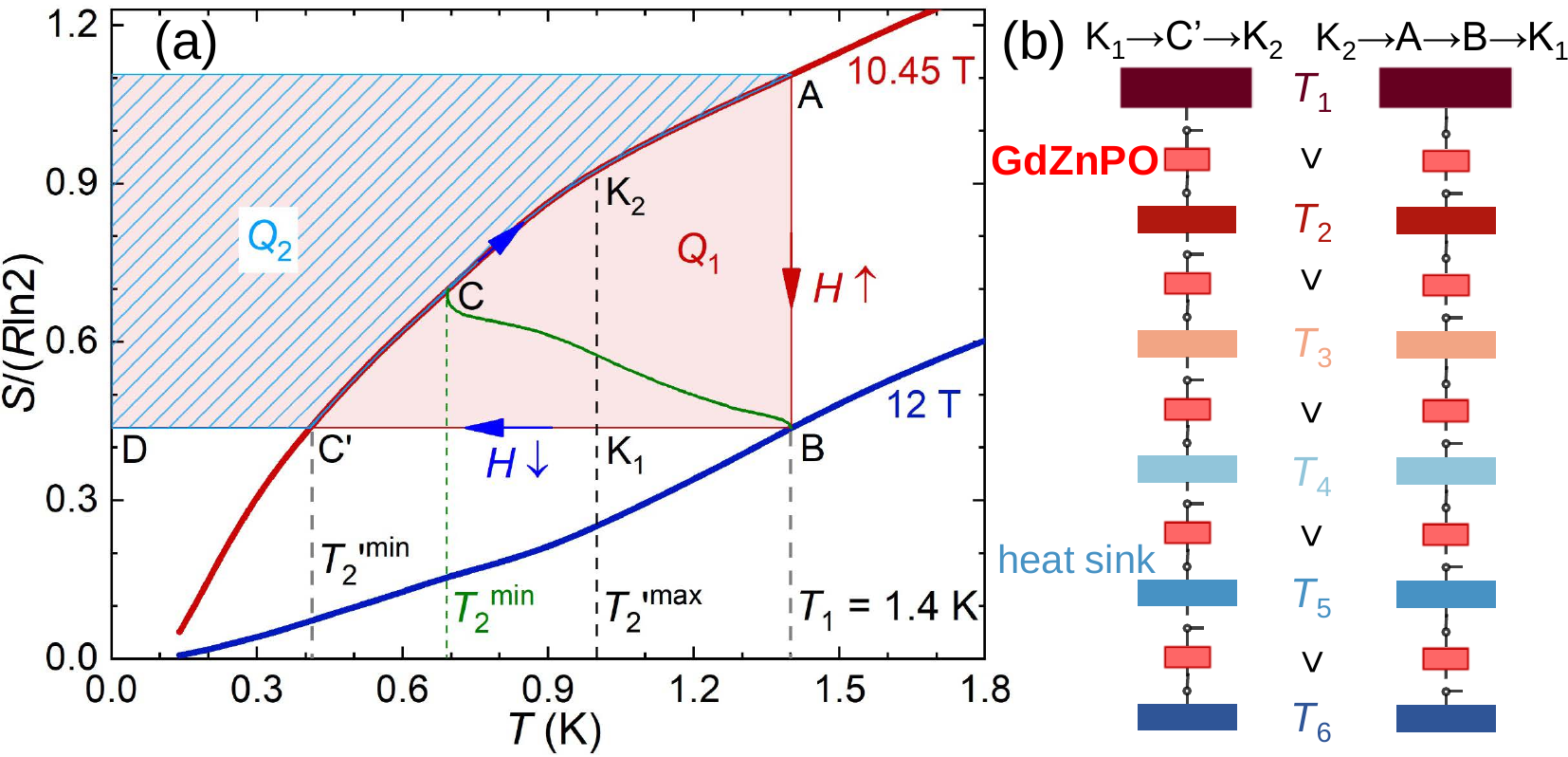}
  \begin{justify}
  \caption{Magnetic refrigeration cycle. a) Entropy ($S$) per mole of GdZnPO, integrated from the specific heat (Figure S5, Supporting Information). The solid olive line shows the B$\to$C route, $S$($T_2$, $H$), where $T_2$ and $H$ are measured by maintaining $T_1$ = 1.4 K, as shown in Figure~\ref{fig3}. b) Schematic blueprint of using multiple GdZnPO samples for magnetic cooling cycles.}
  \label{fig4}
  \end{justify}
\end{center}
\end{figure}

\begin{justify}
GdZnPO exhibits both a giant residual specific heat ($C_0$ $\sim$ 1.2 JK$^{-1}$mol$^{-1}$)~\cite{PhysRevLett.133.236704} and a giant MCE ($\it{\Gamma}_\mathrm{m}$ $\sim$ 4.44 T$^{-1}$) near $H_\mathrm{c}$, indicating high cooling power even at extremely low temperatures. The cooling power, $P_\mathrm{C}$ = $C_\mathrm{tot}T\it{\Gamma}_\mathrm{m}v_H$, was evaluated at $T$ = 0.2 K and $v_H$ = 0.3 Tmin$^{-1}$, yielding $P_\mathrm{C}$ $\sim$ 2.7, 5.0, 5.1, and 1.5 mWmol$^{-1}$ at $\mu_0H$ = 10.8 (where $\it{\Gamma}_\mathrm{m}$ is low), 11, 11.5, and 11.8 T (where $C_\mathrm{tot}$ is small), respectively (Table S1, Supporting Information). This suggests that approximately 1/5 mol, or 50 g, or 8 cm$^3$ of GdZnPO can generate a cooling power of $\sim$1 mW, comparable to the performance of a good $^3$He-$^4$He dilution refrigerator at low temperatures ($\sim$0.2 K).

The entropy was integrated from the measured specific heat~\cite{PhysRevLett.133.236704}, with its reliability discussed in the Supporting Information. The magnetic refrigeration cycle of GdZnPO in the $S$-$T$ space at $T_1$ = 1.4 K is shown in Figure~\ref{fig4}a. In the isothermal process A$\to$B, the applied field drives the system from the SSL to the FM state below $T^*$, resulting in an entropy decrease and heat release, $Q_1$ = $T_1(S_\mathrm{A}-S_\mathrm{B})$ (Figure~\ref{fig4}a). The B$\to$C$'$ process corresponds to an adiabatic demagnetization refrigeration process, with the GdZnPO spin system expected to reach its lowest temperature, $T_2'$$^\mathrm{min}$. In our experiments, the system actually underwent a quasi-adiabatic demagnetization refrigeration process that approximately followed B$\to$C (solid olive line in Figure~\ref{fig4}a), resulting in a slightly higher $T_2$$^\mathrm{min}$. The C$'$$\to$A process is a constant-field relaxation process, with a decalescence contribution, $Q_2$ = $\int_{\mathrm{C'}}^{\mathrm{A}}T\mathrm{d}S(T,H\equiv H_\mathrm{end})$. We obtained a magnetic cooling efficiency of $Q_2$/$Q_1$ = 0.60(2) for GdZnPO at $T_1$ = 0.7-1.8 K (Table S1, Supporting Information).

As shown in Figure~\ref{fig4}a, the decalescence under actual quasi-adiabatic conditions, $Q_2^\mathrm{a}$ = $\int_{\mathrm{B}}^{\mathrm{C}}T\mathrm{d}S+\int_{\mathrm{C}}^{\mathrm{A}}T\mathrm{d}S$, is clearly larger than that under ideal adiabatic conditions, $Q_2$ = $\int_{\mathrm{B}}^{\mathrm{C'}}T\mathrm{d}S+\int_{\mathrm{C'}}^{\mathrm{A}}T\mathrm{d}S$ = $\int_{\mathrm{C'}}^{\mathrm{A}}T\mathrm{d}S$, due to the additional area enclosed by path BC'CB. Consequently, the magnetic cooling efficiency under actual quasi-adiabatic conditions, $Q_2^\mathrm{a}$/$Q_1$, exceeds the ideal adiabatic efficiency, $Q_2$/$Q_1$. This arises because the heat leakage increases the minimum achievable temperature, i.e., $T_2^\mathrm{min}$ $>$ $T_2'$$^\mathrm{min}$. In the specific case shown in Figure~\ref{fig4}a for $T_1$ = 1.4 K, we obtain $Q_1$ = $T_1(S_\mathrm{A}-S_\mathrm{B})$ = 5.406 Jmol$^{-1}$ per cycle, $Q_2$ = 3.157 Jmol$^{-1}$ per cycle, $T_2'$$^\mathrm{min}$ = 0.409 K, $Q_2^\mathrm{a}$ = 3.866 Jmol$^{-1}$ per cycle, and $T_2^\mathrm{min}$ = 0.693 K. As a result, we find $Q_2^\mathrm{a}$/$Q_1$ ($\sim$0.72) $>$ $Q_2$/$Q_1$ ($\sim$0.58) and $T_2^\mathrm{min}$/$T_1$ ($\sim$0.50) $>$ $T_2'$$^\mathrm{min}$/$T_1$ ($\sim$0.29). Although the actual quasi-adiabatic efficiency $Q_2^\mathrm{a}$/$Q_1$ can be higher than the ideal adiabatic efficiency $Q_2$/$Q_1$, the latter is independent of specific experimental conditions and thus better reflects the intrinsic cooling efficiency of a material.

The GdZnPO spin system exhibits no observable magnetic hysteresis down to at least 0.05 K ($\ll |\theta_\mathrm{w}|$, Figure S2, Supporting Information), consistent with the theoretical expectation of the classical SSL ansatz for the easy-plane $J_1$-$J_2$ honeycomb-lattice model at zero temperature~\cite{PhysRevB.106.035113}. Consequently, magnetic dissipation is negligible, and the system entropy is expected to be fully recoverable after a cooling cycle.

\end{justify}

\section{Discussion}

\begin{table}
	\caption{Volumetric entropy capacity ($S_\mathrm{V}$ in JK$^{-1}$cm$^{-3}$), maximum Gr\"{u}neisen parameter ($\it{\Gamma}_\mathrm{m}$ in T$^{-1}$), normalized ratio ($\it{\Gamma}^\mathrm{norm}$ = $\mu_0H\it{\Gamma}_\mathrm{m}$), magnetic cooling efficiency ($Q_2$/$Q_1$), and low-$T$ operational field range ($\mu_0H_\mathrm{sta}$$-$$\mu_0H_\mathrm{end}$ in T) of GdZnPO, compared to other typical magnetocaloric materials.}
	\begin{center}
		\begin{tabularx}{18cm}{>{\centering\arraybackslash}p{4.5cm}||Y|Y|Y|Y|Y}
			\hline
			\hline
			compounds & $S_\mathrm{V}$ & $\it{\Gamma}_\mathrm{m}$ & $\it{\Gamma}^\mathrm{norm}$ & $Q_2$/$Q_1$ & $\mu_0H_\mathrm{sta}$$-$$\mu_0H_\mathrm{end}$ \\\hline
            \textbf{GdZnPO} (this work) & \textbf{0.424} & \textbf{4.44} & \textbf{51} & \textbf{0.60(2)} & \textbf{12$-$11.2}\\
            Na$_2$BaCo(PO$_4$)$_2$~\cite{xiang2024giant} & 0.056 & 3.8 & 6.7 & - & 4$-$1.4\\
            Cs$_2$CuCl$_4$~\cite{pssb.201200794} & 0.043 & 3.4 & 29 & 0.57 & 10$-$8.5\\
            NaYbP$_2$O$_7$~\cite{PhysRevApplied.20.014013} & 0.064 & 3.2 & 0.9 & - & 5$-$0\\
            Cu(NO$_3$)$_2\cdot$2.5H$_2$O~\cite{Vantol1973Specific} & 0.059 & 2.7 & 12 & - & 5$-$2.9\\
            KBaGd(BO$_3$)$_2$~\cite{PhysRevB.107.125126,PhysRevB.107.104402,xiang2023dipolarspinliquidending} & 0.192 & 1.4-3.1 & 0.8-2.7 & - & 6$-$0.75\\
            KBaYb(BO$_3$)$_2$~\cite{tokiwa2021frustrated} & 0.064 & 2.4 & 1.2 & - & 5$-$0\\
            abbreviated to CuP~\cite{wolf2011magnetocaloric} & 0.028 & 1.9 & 8.3 & 0.26 & 7$-$4\\
            EuB$_4$O$_7$~\cite{wang2024giant} & 0.285 & 1.7 & 1.0 & - & 4$-$0\\
            NaYbGeO$_4$~\cite{PhysRevB.108.224415} & 0.101 & 1.7 & 0.9 & - & 5$-$0\\
            YbPt$_2$Sn~\cite{jang2015large} & 0.127 & 1.6 & 0.8 & - & 6$-$0\\
            YbNi$_4$Mg~\cite{zhang2024ybni4} & 0.111 & 1.4 & 0.7 & - & 3$-$0\\
            Er$_2$Ti$_2$O$_7$~\cite{wolf2016magnetic} & 0.148 & 1.3 & 2.2 & 0.53 & 7$-$1.5\\
            NaGdP$_2$O$_7$~\cite{telang2024adibatic} & 0.209 & 1.2 & 1.0 & - & 5$-$0\\
            EuCl$_2$~\cite{wang2024record} & 0.378 & 1.0 & 0.9 & - & 5$-$0\\
            YbCu$_4$Ni~\cite{10.1063/5.0064355} & 0.114 & 1.0 & 0.8 & - & 10$-$0\\
            Gd$_3$Ga$_5$O$_{12}$~\cite{xiang2024giant} & 0.363 & 0.8 & 1.6 & - & 4$-$0\\
            Na$_2$SrCo(PO$_4$)$_2$~\cite{PhysRevMaterials.8.044409} & 0.059 & 0.8 & 2.0 & - & 6$-$0\\
			\hline
			\hline
		\end{tabularx}
	\end{center}
	\label{tab1}
\end{table}

\begin{justify}
As shown in Figure~\ref{fig2}a, $\it{\Gamma}^\mathrm{norm}$ or $\it{\Gamma}_\mathrm{m}$ peaks near the crossover field $\mu_0H_\mathrm{c}$ $\sim$ 12 T, and remains high up to $\sim$4.5 K. Based on this, we propose using multiple GdZnPO samples for magnetic cooling cycles, directly cooling from a high temperature ($\sim$4.5 K) down to $\sim$36 mK within the same superconducting magnet, as illustrated in Figure~\ref{fig4}b. To ensure a temperature decrease from higher to lower heat sinks, when the GdZnPO sample temperature drops below a set value $T_2'$$^\mathrm{max}$ (i.e., during K$_1$$\to$C$'$$\to$K$_2$, see Figure~\ref{fig4}a), the HS between the higher sink and the sample should be turned off, and the HS between the sample and the lower sink should be turned on, as shown in Figure~\ref{fig1}c. Conversely, when the sample temperature exceeds $T_2'$$^\mathrm{max}$ (i.e., during K$_2$$\to$A$\to$B$\to$K$_1$), the HS between the higher sink and the sample should be turned on, and the HS between the sample and the lower sink should be turned off, as shown in Figure~\ref{fig1}d. To ensure a continuous temperature decrease from higher to lower sinks (i.e., $T_{j+1}$ $<$ $T_j$ for $j$ = 1, 2, ..., see Figure~\ref{fig4}b), the HSs can be activated simultaneously at the K$_1$ and K$_2$ points during the adiabatic demagnetization refrigeration cycles, as shown in Figure~\ref{fig4}.

While paramagnetic salts~\cite{10.1142/S0217979214300175}, spin supersolids~\cite{xiang2024giant}, and heavy fermion metals~\cite{zhang2024ybni4,10.1063/5.0064355} have been widely studied for magnetic cooling, the potential of SSLs remains largely unexplored. Our results demonstrate that the $s$ = 7/2 honeycomb-lattice SSL candidate GdZnPO offers a competitive advantage in magnetic cooling over other compounds (Table~\ref{tab1}). The crossover field, $\mu_0H_\mathrm{c}$ $\sim$ 12 T, arising from the strongly correlated magnetism of GdZnPO, is exceptionally high compared to other compounds. Consequently, efficient adiabatic demagnetization refrigeration cycles using GdZnPO can be performed above $\sim$11 T (see Figure~\ref{fig3}), making it particularly useful when a high magnetic field is necessary. While this high operational field entails higher equipment costs, potentially limiting the practical application of GdZnPO in adiabatic demagnetization refrigeration, other refrigerants (like commercial Gd$_3$Ga$_5$O$_{12}$ ~\cite{xiang2024giant,wang2024giant}) require high starting fields of $\mu_0$$H_\mathrm{sta}$ $\sim$ 3-10 T (Table~\ref{tab1}), necessitating superconducting magnets. While increasing the superconducting magnet's maximum field to 12 T incurs additional costs, the energy consumption per cycle is roughly proportional to the operational field range, $\mu_0$($H_\mathrm{sta}$$-$$H_\mathrm{end}$), which is lowest for GdZnPO (Table~\ref{tab1}). The main superconducting coils can provide a steady $\sim$11 T field in persistent mode, requiring no energy input beyond cryogenic maintenance. The remaining $\sim$0-1 T field can be cycled by secondary coils, as in commercial magnetic refrigerators. These features make GdZnPO a promising candidate for magnetic cooling applications.

Due to the extremely strong neutron absorption by Gd atoms, neutron scattering measurements on GdZnPO remain highly challenging. Nevertheless, several experimental observations support the emergence of a SSL in GdZnPO at low temperatures:

(1) Within the spherical approximation, the generic theory of SSLs predicts a distinctive low-temperature behavior of the magnetic specific heat: $C_\mathrm{m}$ $\sim$ $C_0+C_1T$~\cite{bergman2007order,yao2021generic}. To our knowledge, this behavior has not been reported in other spin systems. In a two-dimensional SSL, $\mathbf{Q}_\mathrm{G}$ fluctuates along a continuous contour in reciprocal space, in stark contrast to conventional magnetic states where $\mathbf{Q}_\mathrm{G}$ takes on discrete values. This fluctuation implies the existence of spin degrees of freedom along the spiral contour---analogous to the translational degrees of freedom in an ideal gas---which leads to a finite residual specific heat $C_0$. $C_0$ originates from zero-energy excitations along the degenerate continuous contour in the classical limit ($s$ $\to$ $\infty$), while the linear term $C_1T$ arises from low-energy excitations off the contour. To our knowledge, this low-$T$ behavior of $C_\mathrm{m}$ $\sim$ $C_0+C_1T$ had not been experimentally observed in any compounds until our previous report on the SSL candidate GdZnPO, where this behavior appears below the crossover field $\mu_0H_\mathrm{c}$ $\sim$ 12 T and down to $\sim$53 mK ($\ll$ $|\theta_\mathrm{w}|$)~\cite{PhysRevLett.133.236704}.

(2) Within the quasi-particle framework, the magnetic thermal conductivity of a SSL is expected to scale with the magnetic specific heat, i.e., $\kappa_{xx}^\mathrm{m}$ $\sim$ $\kappa_0+\kappa_1T$, analogous to $C_\mathrm{m}$ $\sim$ $C_0+C_1T$. Recently, we observed a large thermal conductivity in GdZnPO that closely follows the expected form $\kappa_{xx}$ = $\kappa_0+\kappa_1T+K_\mathrm{p}T^3$ below 1 K and up to the crossover field $\mu_0H_\mathrm{c}$ $\sim$ 12 T, where the $K_\mathrm{p}T^3$ term arises from phonons~\cite{zhao2025itinerant}. Notably, the observation of a magnetic contribution $\kappa_{xx}^\mathrm{m}$ $\sim$ $\kappa_0+\kappa_1T$ is highly distinctive and, to our knowledge, has not been reported in any other magnetic compound.

(3) Low-temperature magnetization measurements clearly reveal easy-plane anisotropy in GdZnPO~\cite{PhysRevLett.133.236704}. Taking into account the crystal structure and magnetization data, the spin system is well described by the easy-plane $J_1$-$J_2$ honeycomb-lattice model~\cite{PhysRevLett.133.236704}. Theoretically, this model stabilizes a SSL over a broad parameter range, $\mid$$J_2/J_1$$\mid$ $>$ 1/6, and the SSL persists down to very low temperatures~\cite{PhysRevResearch.4.013121}. Moreover, the experimentally determined Hamiltonian yields a crossover field of $\mu_0H_\mathrm{c}$ = $s[2D+3J_1+9J_2+J_1^2/(4J_2)]/(g\mu_\mathrm{B})$ $\sim$ 12 T and a Curie-Weiss temperature of $\theta_\mathrm{w}$ = $-s(s+1)(J_1+2J_2)$ $\sim$ $-$12 K, both of which are in good agreement with experimental observations on GdZnPO (see Figure 2)~\cite{PhysRevLett.133.236704}.

(4) Within the SSL ansatz on the honeycomb lattice~\cite{PhysRevB.106.035113}, the zero-temperature magnetic susceptibility is theoretically predicted to remain constant up to $\mu_0H_\mathrm{c}$. Specifically, the calculated parallel susceptibility is given by $\chi_\mathrm{cal}^{\parallel}$ = $\mu_0N_\mathrm{A}g^2\mu_\mathrm{B}^2/[2D+3J_1+9J_2+J_1^2/(4J_2)]$+$\chi_\mathrm{vv}^\parallel$, which yields a value of $\sim$4.4 cm$^3$mol$^{-1}$ for GdZnPO, in good agreement with experimental results down to 50 mK ($\sim$0.4\%$\mid$$\theta_\mathrm{w}$$\mid$, Figure S3, Supporting Information)~\cite{PhysRevLett.133.236704}.

(5) In this work, the giant magnetocaloric effect observed in GdZnPO is well captured by the experimentally determined easy-plane $J_1$-$J_2$ honeycomb-lattice spin Hamiltonian~\cite{PhysRevLett.133.236704}, without the need for adjustable microscopic interaction parameters (see Figure~\ref{fig2})---further supporting the presence of a SSL.

Interlayer couplings can effectively increase the system's dimensionality from two to three. Both this dimensional crossover and the spatial anisotropy of the honeycomb lattice tend to relieve magnetic frustration, thereby destabilizing the SSL at low temperatures. For example, the honeycomb-lattice compound FeCl$_3$ exhibits a clear instability of the SSL at $\sim$5 K, likely due to significant interlayer interactions~\cite{PhysRevLett.128.227201}. In contrast, interlayer couplings in GdZnPO are negligible, owing to the highly localized nature of the Gd$^{3+}$ 4$f$ electrons and the large interlayer spacing of $c$/3 $\sim$ 10.2 \AA, where magnetic layers are separated by nonmagnetic ZnP bilayers~\cite{PhysRevLett.133.236704}. Furthermore, the spatial isotropy of the honeycomb lattice in GdZnPO is protected by the $R\overline{3}m$ space group symmetry. As a result, measurements on GdZnPO show no clear indication of instability from the SSL into a conventionally ordered state below $\mu_0H_\mathrm{c}$, down to at least $\sim$50 mK ($\sim$0.4\%$\mid$$\theta_\mathrm{w}$$\mid$).

In the frustrated honeycomb-lattice antiferromagnet GdZnPO, the field-induced crossover from the non-degenerate FM state ($H$ $\geq$ $H_\mathrm{c}$) to the putative degenerate SSL ($H$ $<$ $H_\mathrm{c}$) leads to a significant increase in entropy, decalescence, and a giant MCE, near the crossover field $H_\mathrm{c}$ at low temperatures. We demonstrate that GdZnPO is an efficient material for magnetic cooling, operating effectively within a narrow field range ($\sim$11-12 T, near $\mu_0H_\mathrm{c}$), with high cooling power and excellent performance across a wide temperature range ($\sim$0.036-4.5 K). Our findings not only highlight the potential of GdZnPO for magnetic cooling applications but also open avenues for further research into other SSL candidates for similar purposes.
\end{justify}

\section{Experimental Section}

\begin{justify}
Samples: Single crystals of GdZnPO were prepared using the two-step method~\cite{PhysRevLett.133.236704}. The resulting crystals exhibit generally high quality, as evidenced by sharp and clear Laue diffraction patterns (Figure S3, Supporting Information), and no significant sample dependence of thermodynamic properties has been observed in our recent studies~\cite{zhao2025itinerant}. In general, finite-size effects on thermodynamic properties become negligible when the system exceeds $\sim$1000 spins along each dimension. Consequently, macroscopic structural defects such as grain boundaries have minimal impact on thermodynamic measurements. Therefore, the magnetocaloric effect ($\it{\Gamma}_\mathrm{m}$ = $-(\mathrm{d}M/\mathrm{d}T)/C_\mathrm{p}$) and the associated magnetic cooling performance---both governed by thermodynamic properties---are expected to be largely insensitive to structural imperfections in the case of GdZnPO.

Experimental Methods: Twelve $ab$-plane single crystals (total mass 5.37 mg), aligned with the $c$ axis along the applied magnetic field, were used to measure MCE and conduct quasi-adiabatic demagnetization refrigeration in a $^3$He-$^4$He dilution refrigerator. As shown in the inset of Figure~\ref{fig2}a, a weak thermal link between the sample and the Cu heat sink was established using a thin constantan wire. The MCE was measured using the high-resolution alternating-field technique under quasi-adiabatic conditions~\cite{tokiwa2011high,PhysRevX.10.011007,li2023frustrated}, with a frequency of $\nu$ = 0.005 or 0.01 Hz. Specific heat and magnetization of GdZnPO at $T$ = 0.05-1.8 K and $\mu_0H$ = 0-12 T were measured in the dilution refrigerator~\cite{PhysRevLett.133.236704,PhysRevMaterials.8.074410,zhao2024quantum,shimizu2021development}. Both the MCE and the quasi-adiabatic demagnetization refrigeration results are highly  reproducible (Figure S2, Supporting Information). Further details on the experimental procedures are provided in the Supporting Information.

Numerical Methods: Standard MC simulations of the thermodynamic quantities (Figure~\ref{fig2}b) were performed on a 2$\times$72$^2$ cluster with periodic boundary conditions, using the previously determined Hamiltonian of GdZnPO without parameter tuning~\cite{PhysRevLett.133.236704}. 5,000 MC steps were conducted at each of 200 temperatures, gradually annealing from 50 K to 0.05 K, with 1,000 steps allocated for thermalization. The calculated quantities were averaged over four independent samples. The MC algorithm relies on repeated random sampling, which inherently introduces noise in the calculated $\it{\Gamma}_\mathrm{m}$, especially at low temperatures. This noise is random in nature rather than oscillatory, as its pattern varies noticeably at a different temperature (see Figure~\ref{fig2}b).

Statistical Analysis: All statistical analyses were performed using OriginPro 2018.

\end{justify}

\medskip
\textbf{Supporting Information} \par 
Supporting Information is available from the Wiley Online Library or from the authors.

\medskip
\textbf{Data Availability} \par
The datasets generated during and/or analyzed during the current study are available from the corresponding author on reasonable request.

\medskip
\textbf{Acknowledgements} \par 

We gratefully acknowledge Wei Li, Shang Gao, and Shiliang Li for helpful discussion. This work was supported by the National Key R\&D Program of China (Grants No. 2024YFA1613100 and No. 2023YFA1406500), the National Natural Science Foundation of China (Grant No. 12274153), and the Fundamental Research Funds for the Central Universities (No. HUST: 2020kfyXJJS054).

\medskip
\textbf{Conflict of Interest} \par
The authors declare no conflict of interest.

\medskip

%



\begin{figure}
\textbf{Table of Contents}\\
\medskip
  \includegraphics{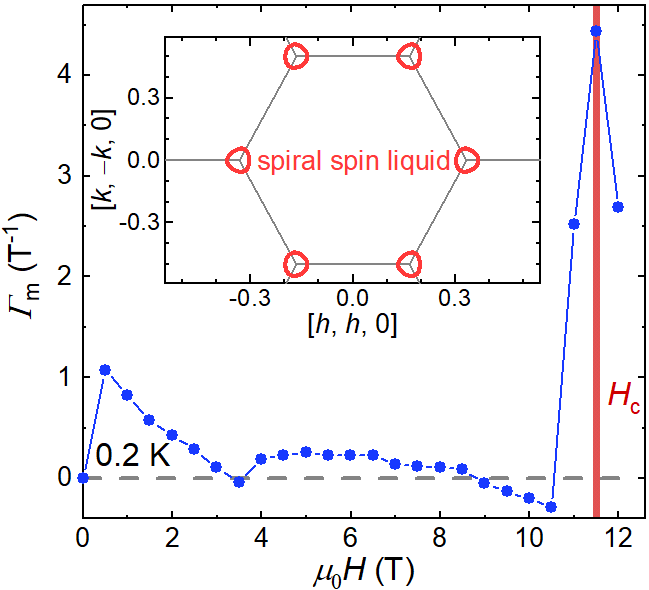}
  \medskip
  \begin{justify}
  \caption*{The frustrated honeycomb-lattice single-crystal GdZnPO, a spiral spin-liquid (SSL) candidate, exhibits a giant magnetocaloric effect (with a large Gr\"{u}neisen parameter $\it{\Gamma}_\mathrm{m}$ = 4.4 T$^{-1}$) and exceptional cooling performance near the crossover magnetic field ($H_\mathrm{c}$), surpassing other known magnetocaloric materials. These findings underscore GdZnPO's promise for magnetic cooling and open avenues for further research into SSL candidates for similar applications.}
  \end{justify}
\end{figure}

\end{document}